\pdfoutput=1
\documentclass[11pt]{article}
\usepackage[preprint]{acl}
\usepackage{times}
\usepackage{latexsym}
\usepackage[T1]{fontenc}
\usepackage[utf8]{inputenc}
\usepackage{microtype}
\usepackage{inconsolata}
\makeatletter
\@namedef{T1/zi4/m/it}{<->ssub*zi4/m/n}
\@namedef{T1/zi4/b/it}{<->ssub*zi4/b/n}
\@namedef{T1/zi4/bx/it}{<->ssub*zi4/b/n}
\@namedef{T1/zi4/m/sl}{<->ssub*zi4/m/n}
\@namedef{T1/zi4/b/sl}{<->ssub*zi4/b/n}
\@namedef{T1/zi4/bx/sl}{<->ssub*zi4/b/n}
\makeatother
\usepackage{graphicx}
\usepackage{algorithm}
\usepackage{algpseudocode}
\usepackage{shellesc}
\usepackage{hyperref}
\usepackage{url}
\usepackage{booktabs}
\usepackage{caption}
\usepackage{array}
\usepackage{adjustbox}
\usepackage{longtable}
\usepackage{lineno}
\usepackage{listings}
\usepackage{tabularx}
\usepackage{adjustbox}
\usepackage{minted}
\usepackage{pgfplots}
\usepackage{pgfplotstable}
\usepackage{amsmath}
\usepackage{multirow}
\usepackage{amsfonts}
\usepackage{float}
\pgfplotsset{compat=1.18}

\title{Enhancing Financial RAG with Agentic AI and Multi-HyDE: A Novel Approach to Knowledge Retrieval and Hallucination Reduction}

\author{
 \textbf{Akshay Govind Srinivasan\textsuperscript{*}\textsuperscript{1}},
 \textbf{Ryan Jacob George\textsuperscript{*}\textsuperscript{1}},
 \textbf{Jayden Koshy Joe\textsuperscript{*}\textsuperscript{1}},
 \textbf{Hrushikesh Kant\textsuperscript{*}\textsuperscript{1}},
 \\
 \textbf{Harshith M R\textsuperscript{*}\textsuperscript{1}},
 \textbf{Sachin Sundar\textsuperscript{*}\textsuperscript{1}},
 \textbf{Sudharshan Suresh\textsuperscript{*}\textsuperscript{1}},
 \textbf{Rahul Vimalkanth\textsuperscript{*}\textsuperscript{1}},
 \textbf{Vijayavallabh\textsuperscript{*}\textsuperscript{1}}
\\
\\
 \textsuperscript{1}Indian Institute of Technology Madras
}

\begin{document}
\maketitle
\let\thefootnote\relax\footnotetext{*These authors contributed equally to this work}

\begin{abstract}
Accurate and reliable knowledge retrieval is vital for financial question-answering, where continually updated data sources and complex, high-stakes contexts demand precision. Traditional retrieval systems rely on a single database and retriever, but financial applications require more sophisticated approaches to handle intricate regulatory filings, market analyses, and extensive multi-year reports. We introduce a framework for financial Retrieval Augmented Generation (RAG) that leverages agentic AI and the Multi-HyDE system, an approach that generates multiple, nonequivalent queries to boost the effectiveness and coverage of retrieval from large, structured financial corpora. Our pipeline is optimized for token efficiency and multi-step financial reasoning, and we demonstrate that their combination improves accuracy by 11.2\% and reduces hallucinations by 15\%. Our method is evaluated on standard financial QA benchmarks, showing that integrating domain-specific retrieval mechanisms such as Multi-HyDE with robust toolsets, including keyword and table-based retrieval, significantly enhances both the accuracy and reliability of answers. This research not only delivers a modular, adaptable retrieval framework for finance but also highlights the importance of structured agent workflows and multi-perspective retrieval for trustworthy deployment of AI in high-stakes financial applications.
\end{abstract}

\section{Introduction}
\label{sec:intro}
Large Language Models (LLMs) such as GPT-4 \citep{openai2024gpt4technicalreport}, LLaMA \citep{touvron2023llama}, and PaLM \citep{chowdhery2022palm} have significantly advanced natural language processing, demonstrating strong capabilities in contextual reasoning and few-shot learning. These models are increasingly applied in high-stakes domains, including healthcare diagnostics \citep{singhal2023large}, legal document analysis \citep{henderson2023foundation}, and financial services \citep{wu2023bloomberggpt, li2023finbert}. Their ability to process and generate domain-specific, human-like responses offers clear potential benefits.

However, a persistent limitation of LLMs is \textit{hallucination} - the generation of factually incorrect or fabricated content presented as truth \citep{ji2023survey, huang2023survey}. This limitation poses significant risks in domains where factual accuracy is paramount.
In domains such as finance, where decisions must be based on accurate and verifiable data, hallucinations can lead to significant monetary losses, reputational harm, and regulatory violations.

Retrieval-Augmented Generation (RAG) frameworks \citep{lewis2020retrieval,guu_realm_2020} address this issue by grounding LLM outputs in external knowledge sources. Conventional RAG pipelines use a retriever to fetch relevant document chunks from a database based on semantic similarity between vector embeddings \citep{karpukhin2020dense, xiong2020approximate}. Improvements in retrieval have come from better embedding methods \citep{reimers2019sentence, gao2021simcse}, hybrid dense-sparse strategies, and hierarchical retrieval \citep{khattab2020colbert, zhang2022designing}.

One particularly effective method for improving retrieval is Hypothetical Document Embeddings (HyDE) \citep{gao2023precisezeroshotdenseretrieval}, where an LLM first generates a synthetic “hypothetical” answer to a query, embeds it, and then retrieves real documents most similar to that synthetic answer. This approach improves alignment between queries and relevant passages, especially in cases where the original query is underspecified or phrased differently than the source content.

Recent work in \textit{Agentic RAG} \citep{schick2023toolformer, qin2023tool, yao2022react, liu2023toolllm} extends the static “retrieve-then-generate” pipeline into a dynamic decision-making process. Here, the LLM acts as an orchestrator, capable of decomposing complex queries, selecting appropriate tools or retrieval strategies, performing multi-hop searches, and verifying intermediate results before generating a final answer. Such systems have shown particular promise in domains requiring multi-step reasoning and evidence verification, making them well-suited for financial question answering, where queries may range from straightforward fact lookups to multi-document analyses \citep{wang2025finsagemultiaspectragfinancial}.

Financial QA systems must process vast repositories of unstructured data, including annual reports, regulatory filings, earnings call transcripts, and market analyses \citep{wu2023bloomberggpt, li2023finbert}. The retrieval strategy must be both accurate and efficient, as inadequate retrieval can lead to irrelevant or misleading context being passed to the LLM. This is especially problematic for multi-hop queries, where context mismanagement or excessive token usage can degrade performance despite the availability of long-context models. Methods that involve processing information in the data stores into structures like graphs result in increased upfront token costs, albeit with better performance.
To address these challenges in financial question answering, we present the following contributions:
\begin{itemize}
    \item \textbf{Multi-HyDE}: A retrieval mechanism that utilizes multi-perspective hypothetical documents bringing an improvement in retrieval accuracy without an increase in token costs over HyDE \cite{gao2023precisezeroshotdenseretrieval}
    \item A combination of dense and sparse retrieval strategies to maintain performance on vector stores with over 500,000 tokens.
    \item An Agentic system that is capable of handling both straightforward queries and ones requiring planning, multi-hop retrieval, tool calling and verification.
\end{itemize}
Details of our system have been discussed in detail in Section 3. Details about the evaluation set up have been discussed in Section 4.

\section{Related works}
\label{sec:related-works}
\subsection{Retrieval Methods}
The efficacy of Retrieval-Augmented Generation (RAG) systems fundamentally depends on the quality of their retrieval component \citep{lewis2020retrieval,guu_realm_2020}. Traditional RAG implementations employ semantic similarity search over vector databases, but this approach often suffers from a semantic mismatch between concise queries and the verbose, context-rich nature of source documents \citep{langchain2023querytransformations}. To address this, recent research has focused on enhancements in three main categories: pre-retrieval query transformations, hybrid retrieval strategies, and post-retrieval processing.

\paragraph{Pre-retrieval Query Transformation}
Pre-retrieval Query Transformation bridges the semantic gap through sophisticated query manipulation. A seminal advancement is Hypothetical Document Embeddings (HyDE), which uses a language model to generate a ``pseudo-document'' representing an ideal answer. The embedding of this richer document is then used for retrieval, shifting the paradigm from a query-to-document to a more effective answer-to-answer similarity search \citep{gao2023precisezeroshotdenseretrieval}. Parallel to this, multi-query strategies improve recall by generating several variations of a user's query to capture different facets of the information need \citep{langchain2023querytransformations}. However, generating merely similar queries can sometimes degrade precision \citep{eibich2024aragogadvancedragoutput}. Recent advances include DMQR-RAG (Diverse Multi-Query Rewriting) \cite{li2024dmqrragdiversemultiqueryrewriting}, which operates at different information granularity levels, and MUGI (Multi-Text Generation Integration) \cite{zhang2024exploringbestpracticesquery}, a training-free approach that generates multiple pseudo-references to enhance both sparse and dense retrieval. While these approaches improve retrieval, they fundamentally rely on query similarity rather than the complementary diversity we propose.

\paragraph{Hybrid Retrieval Strategies}
Hybrid Retrieval Strategies combine sparse and dense methods to leverage both keyword matching and semantic similarity. Dense retrieval excels at capturing semantic connections but can struggle with exact term matching, while sparse methods like BM25 provide precise keyword matching. In the context of large, structured financial reports, methods relying on vector similarity alone often fail to retrieve all relevant information and struggle to disambiguate semantically similar sections that differ only in critical numerical or temporal details. Our framework explicitly integrates Multi-HyDE with BM25 in a unified pipeline optimized for these documents, improving coverage and disambiguation.

\paragraph{Post-retrieval Processing}
Post-retrieval Processing has evolved beyond simple re-ranking to incorporate sophisticated correction mechanisms. For instance, CRAG introduces a retrieval evaluator that assesses document quality and triggers corrective actions, like web searches, when quality is insufficient \citep{yan2024correctiveretrievalaugmentedgeneration}. Self-RAG trains language models to adaptively retrieve passages and self-critique through generated reflection tokens \citep{asai2023selfraglearningretrievegenerate}. MAIN-RAG proposes a multi-agent filtering framework where agents collaboratively score retrieved documents \citep{chang2024mainragmultiagentfilteringretrievalaugmented}. While promising, these systems introduce computational overhead to fix retrieval issues. Our approach therefore also emphasizes improving retrieval quality from the outset to reduce the need for extensive correction.

Our Multi-HyDE generates multiple non-equivalent but contextually related queries. Unlike methods that create semantically similar queries, our approach creates distinct but complementary information needs—for instance, generating separate queries about a company's fraud investigations and its criminal cases that might be answered within the same document context.

\subsection{Agentic RAG}
The static retrieve and generate workflow of traditional RAG is insufficient for complex queries that require multi-step reasoning and dynamic information gathering. This has spurred the development of Agentic RAG, which embeds autonomous agents into RAG pipelines to create dynamic problem-solving systems.

\paragraph{Finite State Machine Approaches}
Finite State Machine Approaches structure agentic workflows through formal state management. StateFlow models language model workflows as finite state machines, distinguishing between ``process grounding'' via states and ``sub-task solving'' through actions \citep{wu2024stateflowenhancingllmtasksolving}. This approach has achieved 13-28\% higher success rates than ReAct on benchmarks while reducing costs by 3-5$\times$. Our work extends this paradigm. In contrast to prior work applying state management primarily to retrieval and generation, we extend it to govern all tool calls issued by the language model, enabling coherent reasoning across multiple modalities.

\paragraph{Multi-Agent Architectures}
Multi-Agent Architectures coordinate specialized agents for complex tasks. MAIN-RAG exemplifies this with its multi-agent filtering system \citep{chang2024mainragmultiagentfilteringretrievalaugmented}. However, such multi-agent systems can suffer from increased complexity and failure points.

\subsection{RAG in Finance}
Financial RAG systems face unique challenges due to complexity, precision, and regulation. These include handling 100+ page multi-year reports, disambiguating semantically similar sections, and managing numerical precision where subtle differences have significant implications. Specialized Financial Platforms have emerged to address these challenges.

\paragraph{Specialized Financial Platforms}
FinRobot provides a four-layer architecture with Financial AI Agents and Multi-source Foundation Models \citep{yang2024finrobotopensourceaiagent}. While comprehensive, it lacks the specialized retrieval innovations for financial document disambiguation that our Multi-HyDE approach directly addresses.

FinSage focuses on regulatory compliance through a multi-aspect RAG framework, achieving 92.51\% recall and a 24.06\% accuracy improvement over baselines \citep{wang2025finsagemultiaspectragfinancial}. However, FinSage relies on standard HyDE rather than our multi-perspective approach and uses curated questions instead of a comprehensive benchmark evaluation.

\paragraph{Financial Knowledge Graph Integration}
Financial Knowledge Graph Integration handles complex relationships through structured representations. While promising, knowledge graph approaches require significant upfront processing costs and may not adapt well to rapidly changing financial information. Our approach offers greater flexibility and lower preprocessing overhead while achieving comparable performance through retrieval optimization.

\paragraph{Evaluation Challenges}
Evaluation challenges in finance are complicated by the need for numerical precision. FinanceBench reveals that GPT-4-Turbo with retrieval systems incorrectly answers or refuses 81\% of its questions \citep{islam2023financebenchnewbenchmarkfinancial}. ConvFinQA highlights challenges in conversational queries requiring extensive calculations \citep{chen2022convfinqaexploringchainnumerical}. These issues suggest that many existing systems may report inflated performance due to flawed evaluation methodologies. Our emphasis on human evaluation provides more accurate assessments for high-stakes applications. Our framework's modular design and reliability-focused architecture directly address enterprise deployment concerns often overlooked in academic research, demonstrating that retrieval optimization may provide greater returns than developing domain-specific language models alone.

In summary, existing RAG systems face key challenges including retrieval issues with semantic ambiguity in complex financial texts, limited capacity for multi-step reasoning and calculations, and inefficiencies due to complex architectures and flawed evaluations. Our framework addresses these by using Multi-HyDE with hybrid BM25 to improve retrieval accuracy and disambiguation, integrating an agentic tool usage system governed by unified state management for advanced reasoning, and reducing overhead by avoiding heavy knowledge graphs while relying on human evaluation for realistic performance assessment. This approach enhances retrieval reliability, reasoning capabilities, and system efficiency for financial RAG applications.

\section{Methodology}

To address the challenges outlined in Sections~\ref{sec:intro} and~\ref{sec:related-works}, 
we propose a retrieval-augmented generation (RAG) pipeline with the following key components:

\begin{itemize}
    \item \textbf{Multi-HyDE:} A multi-hypothesis document expansion module that generates several hypothetical documents based on diverse variants of the input query. These documents are then used to retrieve semantically relevant content from the vector store.
    
    \item \textbf{Keyword-based Retrieval:} An auxiliary keyword-based retriever (e.g., BM25) designed to enhance retrieval performance for structured data such as tables, as well as for semantically similar documents (e.g., annual reports across different years).
    
    \item \textbf{Agentic Pipeline:} A multi-stage reasoning and retrieval process comprising:
    \begin{enumerate}
        \item \emph{Query Clarification:} The system first seeks to clarify the user’s question, either through direct interaction with the user or by leveraging web search.
        \item \emph{Initial Retrieval:} The clarified query is used to perform retrieval from the vector store using the components described above.
        \item \emph{Iterative Refinement:} If the retrieved content is unsatisfactory, the system formulates a retrieval plan. This includes the ability to perform multi-hop retrievals, invoke external tools, and decompose the query into sub-queries.
        \item \emph{Final Response:} Once the retrieved evidence is deemed sufficient, the system synthesizes and delivers the final answer to the user.
    \end{enumerate}
\end{itemize}

This integrated design allows the pipeline to combine the semantic strengths of vector-based retrieval with the precision of keyword-based methods, while also enabling dynamic reasoning for complex, multi-step information needs.

\subsection{Multiple Hypothetical Dynamic Embeddings (Multi-HyDE)}
\label{subsec:multi-hyde}
For our main retrieval tool, we employ a combination of multi-query based retrieval \citep{eibich2024aragogadvancedragoutput} and HyDE \citep{gao2023precisezeroshotdenseretrieval}, which we call \textit{Multi-HyDE}, along with BM25 based retrieval for tables and a re-ranker.

\paragraph{HyDE} \citet{gao2023precisezeroshotdenseretrieval} employ a generator $g$ to create multiple hypothetical documents from a query $q$ and retrieves real documents $d_i$ from the dataset $\mathcal{D}$ that are similar to the hypothetical ones. $N$ documents are sampled from $g$. An embedding model $f$ is used to generate ``hypothetical document embeddings'' $\hat{v}$ for a query $q$ as depicted in Equation \ref{eqn:HyDE}.
\begin{equation}
\label{eqn:HyDE}
\hat{v} = \frac{1}{N} \sum_{\hat{d_i}\sim g(q)} f(\hat{d_i})
\end{equation}

\paragraph{Multi-HyDE}Multi-query approaches usually generate similar queries to the user's, but this has been shown to reduce retrieval precision \citep{eibich2024aragogadvancedragoutput}. Our approach instead uses an LLM $g_q$ to generate queries $[q_1, q_2, ..., q_N]$ that may have answers present in the same context, following which it generates a hypothetical document for each query. These queries may take the form of similar queries, related queries with distinct meanings (such as including a query on fraud by a company A and a query on criminal cases by company A) or it may result in query decomposition. To the best of our knowledge, this particular approach has not been tried before. An embedding model $f$ is used to generate "hypothetical document embeddings" $\hat{v}_{i} \in \mathbb{R}^{\hat d_{embed}}$, as depicted in Algorithm \ref{algo:multi-hyde}. Our retriever $h$ retrieves $k_1$ documents from $\mathcal{D}$, and we further use a reranker to select the top $k_2$ documents.

\begin{algorithm}
\caption{Multi-HyDE Retrieval}
\label{algo:multi-hyde}
\begin{algorithmic}[1]
\Require query $q$, database $\mathcal{D}$, query and document generators $g_q, g$, embedding model $f$, retriever $h$, reranker $r$, hyperparameters $N, k_1, k_2$
\State $[q_1, \dots, q_N] \gets g_q(q)$
\For{each $q_i$ in $[q_1, \dots, q_N]$}
    \State $\hat{v}_i \gets f(g(q_i))$
    \State $S_i \gets h(\hat{v}_i)$
\EndFor
\State $d_{total} \gets concat(S_1, S_2, ..., S_N)$
\State $d_{final} \gets r(d_{total})$
\State \textbf{return} $d_{final}$
\end{algorithmic}
\end{algorithm}

\subsection{Agentic RAG}
\begin{figure*}
    \centering
    \includegraphics[width=1\linewidth]{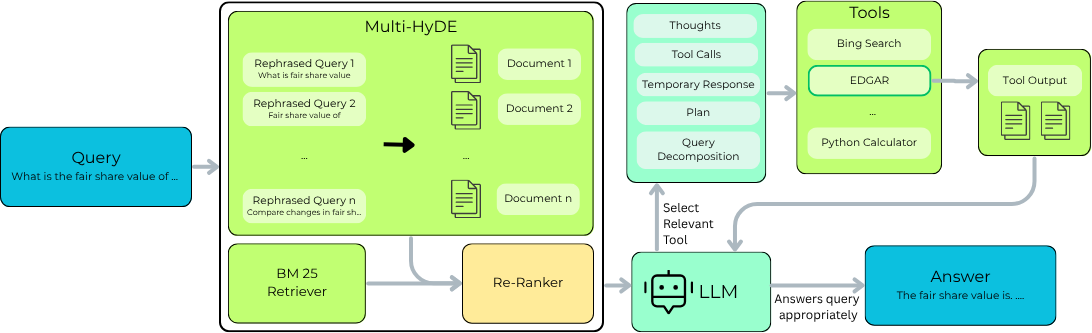}
    \caption{Our proposed agentic framework for financial question-answering.}
    \label{fig:HyFER}
\end{figure*}

To address both simple and complex multi-hop queries, we employ an agentic system (Figure \ref{fig:HyFER}) equipped with several tools, including \texttt{edgar\_tool}, \texttt{Alpha Vantage Exchange Rate}, \texttt{web\_search}, and a Python calculator, as well as a retriever based on Multi-HyDE. Additional tools are listed in Appendix \ref{subsec:toolset}.  

The query processing begins with direct retrieval using Multi-HyDE, ensuring the system remains grounded in explicitly-included sources. Retrieved documents are then passed to the LLM Agent for reasoning and synthesis. If these documents are insufficient to fully answer the query, the LLM dynamically invokes available tools.  

For improved performance, the LLM not only generates tool calls but also produces intermediate reasoning steps, user-facing responses, decomposed sub-queries, and a structured execution plan, inspired by \citet{hao2023reasoninglanguagemodelplanning,radhakrishnan2023questiondecompositionimprovesfaithfulness,zhou2023leasttomostpromptingenablescomplex,wang2023planandsolvepromptingimprovingzeroshot,girhepuje2024regainsenchantintelligenttool}. The full prompt is given in Appendix \ref{sec:metaplanjsonscehma}. Queries are broken down into atomic steps, with each step resolved using the most suitable tool from the current toolset. The LLM evaluates intermediate results at each stage, adapting the plan when necessary to ensure accuracy and grounding.  

This design supports highly dynamic workflows: tools can be added or removed on demand, enabling integration of custom data sources, access to live information, and execution of complex sequential reasoning processes. While standard RAG also grounds responses in retrieved documents, it typically relies on a single retrieval step, leaving the model prone to filling gaps with its latent knowledge if the evidence is incomplete. In contrast, Agentic RAG decomposes queries into atomic steps, validates intermediate results, and dynamically invokes additional tools or retrievals as needed. This iterative, evidence-driven process strengthens fidelity to verifiable sources, reduces hallucination, and produces more reliable answers across diverse and complex query types.

\begin{algorithm}
\caption{Agentic RAG System}
\label{algo:agentic-rag}
\begin{algorithmic}[1]
\Require query $q$, database $\mathcal{D}$, set of tools $T$, LLM agent $A$
\Function{Process\_Query}{$q, \mathcal{D}, T, A$}
    \State $d_{initial} \gets \text{Multi-HyDE}(q, \mathcal{D})$
    \State LLM Agent history $H \gets [q, d_{initial}]$
    \Loop
        \State $A$ analyzes $H$ to determine if the query can be answered
        \If{$A$ determines an answer exists}
            \State Generate final answer from $H$
            \State \textbf{return} Final answer
        \Else
            \State $A$ generates a sub-query $q_{sub}$ and selects a tool $t \in T$
            \State $tool\_output \gets t(q_{sub})$
            \State $H \gets \text{concat}(H, tool\_output)$ \Comment{Add tool's output to the LLM's history}
        \EndIf
    \EndLoop
\EndFunction
\end{algorithmic}
\end{algorithm}

\section{Experimental setup}
\label{experimental-setup}
We ran our experiments using subsets of datasets (selection of subset is described in Appendix \ref{appendix: dataset}) due to limited resources. We employ GPT-4o mini and the Mini-LM reranker for running the pipeline. Additional implementation details are included in Appendix \ref{implementation-details}.
\subsection{Evaluation datasets}
\label{subsec:eval_data}
We use a subset of questions from the FinanceBench \citep{islam2023financebenchnewbenchmarkfinancial} and ConvFinQA \citep{chen2022convfinqaexploringchainnumerical} datasets. From FinanceBench, we have selected from 150 human-annotated examples provided. These examples include evidence designated as ground truth context, with additional justification considered when necessary. Appendix \ref{appendix: dataset} provides details on the subsets used.
\textbf{Furthermore, we ensure that the entire PDF document is added to the vector store in contrast to the ConvFinQA and FinanceBench datasets, which only pass the evidence pages.} We believe this better simulates real-world scenarios. This restricts us from comparing other retrieval methods where only evidence pages are passed as context. We also include a subset of filings and questions from financial-qa-10K\footnote{\url{https://huggingface.co/datasets/virattt/financial-qa-10K}} for a comparative study. An example of a question-answer pair is provided in Appendix \ref{json-example}.

\subsection{Metrics}
We evaluate experiments and optimizations using \textbf{ROUGE,}\footnote{Low ROUGE scores in some experiments are attributed to the fact that the ground truth answers in the dataset consisted of only a single number, whereas large models explained their approaches.} \textbf{Cosine Similarity}, as well as metrics from RAGAS \citep{es2023ragasautomatedevaluationretrieval} and human evaluation. We use RAGAS with GPT-4o mini to calculate \textbf{Factual Correctness} (similar to the F1 Score) and \textbf{Faithfulness}. During human evaluation, accuracy and reliability are measured. The metrics are defined in Appendix \ref{appendix:def-metrics}.

\section{Experimental results and analyses}

\begin{table}[htbp]
    \centering
    \begin{adjustbox}{max width=1\textwidth}
    \begin{tabular}{@{}ccc@{}}
        \toprule
        \textbf{Method} & \textbf{Accuracy (\%)} & \textbf{Reliability (\%)} \\
        \midrule
        Multi-HyDE & 34.4 & 37.91 \\
        Final Pipeline & \textbf{45.6} & \textbf{52.91} \\
        \bottomrule
    \end{tabular}
    \end{adjustbox}
    \caption{Human evaluation on subset of ConvFinQA and FinanceBench.}
    \label{tab: human_eval}
\end{table}

\begin{table*}
    \centering
    \begin{adjustbox}{max width=1\textwidth}
        
    \begin{tabular}{@{}cccccc@{}}
        \toprule
        \textbf{Method} & \textbf{Cosine Similarity} & \textbf{Recall} &\textbf{Factual Correctness} & \textbf{Faithfulness} & \textbf{ROUGE score*}\\
        \midrule
        Multi-HyDE & 0.6269 & \textbf{0.3547} & \textbf{0.3849} & \textbf{0.8404} & \textbf{0.0594} \\
        HyDE & 0.7660 & 0.1154 & 0.2890 & 0.8290 & 0.0498 \\
        CRAG & 0.7939 & 0.1556 & 0.0855 & 0.2521 & 0.0443 \\
        LightRAG & \textbf{0.7999} & 0.0000 & 0.2434 & 0.4629 & 0.1632 \\
        \bottomrule
    \end{tabular}
    \end{adjustbox}
    \caption{Evaluation Metrics for Different Methods on subset of \textbf{ConvFinQA + FinanceBench}.}
    \label{tab: evaluation metrics on ConvFinQA + FinanceBench}
\end{table*}
\begin{table*}
    \centering
    \begin{adjustbox}{max width=1\textwidth}
        
    \begin{tabular}{@{}cccccc@{}}
        \toprule
        \textbf{Method} & \textbf{Cosine Similarity} & \textbf{Recall} &\textbf{Factual Correctness} & \textbf{Faithfulness} & \textbf{ROUGE score}\\
        \midrule

        Multi-HyDE & 0.8976 & 0.8170 & 0.5205 & \textbf{0.9352} & 0.\textbf{4871} \\
        HyDE & 0.8883 & 0.6885 & \textbf{0.5585} & 0.8463 & 0.3726 \\
        CRAG & \textbf{0.9347} & \textbf{0.8500} & 0.4708 & 0.7774 & 0.4290 \\
        LightRAG & 0.7308 & 0.0000 & 0.0368 & 0.4629 & 0.3412 \\

        \bottomrule
    \end{tabular}
    \end{adjustbox}
    \caption{Evaluation Metrics for Different Methods on subset on questions from financial-qa-10K .}
    \label{tab: evaluation metrics on NVIDIA}
\end{table*}
\paragraph{Performance against other methods}
We provide a comparison of our pipeline against a representative method for retrieval optimization (HyDE), graph based knowledge organization (LightRAG) \cite{guo2024lightragsimplefastretrievalaugmented} and post-retrieval corrective measures (CRAG) \cite{yan2024correctiveretrievalaugmentedgeneration}. We include scores for Multi-HyDE with access to tools against these baselines.

Our results show improvement across all measures except Cosine Similarity: we achieve significant improvements in Recall, Facutal accuracy and Faithfulness (See table \ref{tab: evaluation metrics on ConvFinQA + FinanceBench}, \ref{tab: evaluation metrics on NVIDIA}) while having the same token costs involved as HyDE (since both generate the same number of hypothetical documents for a given user query) and avoid the upfront costs associated with graph based methods to create the graph.

Our approach supports dynamic vector stores - documents can be added or removed from the vector store without incurring additional costs, graph based approaches where removing information from the graph would incur some costs.

The results show the advantages of Multi-HyDE in the financial domain. We attribute the improved performance to the fact that financial reports across multiple years could have semantically similar content - pairing a dense retrieval method that identifies relevant information from an increased variety of potential sources and a sparse keyword based retriever to identify structured information improves overall performance by being able to handle more cases than any individual method.

\paragraph{Reliability Considerations: }
In verifying the LLM as a judge procedure utilized by RAGAS, we observe that in numerical examples, the LLM judge might provide incorrect evaluations (see Appendix \ref{appendix:ragas-failure}). Further, cases where the wrong answer is provided confidently has greater chance of adverse impact that the system admitting to not having the exact answer.
To confirm the performance of our proposed pipeline in light of the above challenges, we conduct a human evaluation of the responses with metrics reliabilty (fraction of confidently given answers which are correct rather than hallucinations) and accuracy (fraction of correct answers). Detailed definitions are provided in \ref{appendix:def-metrics}. 

\paragraph{Ablation study: } In Table \ref{tab:bm25_reranker}, we show that Multi-HyDE outperforms regular HyDE. We also perform a comparison between 2 rerankers \texttt{ms-marco-MiniLM-L-6-v2} (Cross Encoder) and \texttt{bge-reranker-v2-m3} (BGE) from huggingface. Though BGE is more performant, it is significantly more resource-intensive and slower. We also show that hybrid retrieval with BM25 clearly outperforms dense retrieval methods for long-document financial data. Tool calling does not improve accuracy, however it provides resiliency when some types of relevant data are not provided.
\begin{table*}
\begin{adjustbox}{max width=1\textwidth}
    \begin{tabular}{@{}cccccc@{}}
    \toprule
    \textbf{Method} & \textbf{Cosine Similarity} & \textbf{Recall} & \textbf{Factual Correctness} & \textbf{Faithfulness} & \textbf{ROUGE score} \\
    \midrule
    1 & 0.8883 & 0.6885 & 0.5585 & 0.8463 & 0.3726 \\
    2 & 0.8932 & 0.7464 & 0.5539 & 0.8837 & 0.3575 \\
    3 & 0.8935 & \textbf{0.8484} & \textbf{0.5868} & 0.8768 & 0.3996 \\
    4 & 0.8976 & 0.8170 & 0.5205 & \textbf{0.9352} & 0.\textbf{4871} \\    
    5 & \textbf{0.9119} & 0.8033 & 0.5172 & 0.8298 & 0.4628 \\
    6 & 0.8935 & \textbf{0.8484} & 0.5867 & 0.8767 & 0.3996 \\
    \bottomrule
    \end{tabular}
\end{adjustbox}
\caption{Effect of BM25, rerankers and tools on recall. (with financial-qa 10k dataset) \\
\\
1. HyDE \\
2. Multi-HyDE + Cross Encoder Reranker \\
3. Multi-HyDE + BM25 + Cross Encoder Reranker}
4. Multi-HyDE + BM25 + BGE Reranker \\
5. Multi-HyDE + BM25 + BGE Reranker without tools \\
6. Multi-HyDE + BM25 + Cross Encoder Reranker without tools \\
\label{tab:bm25_reranker}
\end{table*}

\section{Future work}
    \paragraph{Agents and fine-tuning} Small Language Models finetuned using parameter efficient techniques like LoRA\cite{hu2021loralowrankadaptationlargeb} to be used as individual agents instead of relying on large closed source models, especially for tasks like query re-writing or hypothetical document generation, particularly to suit the language and format used in financial reports.
   
    \paragraph{Better metrics for financial RAG} Currently, LLM-based evaluation often incorrectly evaluates responses, especially when an answer is primarily numeric. Different evaluation systems may help improve this. In addition, a more comprehensive evaluation on complete datasets could be undertaken given more resources.
   
\section{Conclusion}
This research presents a novel approach to financial question answering, addressing key challenges in hallucination reduction and accurate information retrieval from complex financial documents. Our framework introduces Multi-HyDE, an extension of Hypothetical Document Embeddings that leverages multiple non-equivalent queries to enhance retrieval effectiveness. When combined with BM25 for tables and appropriate rerankers, Multi-HyDE demonstrates superior performance in capturing relevant information from financial corpora. Additionally, we developed and evaluated an agentic pipeline offering improved performance, capable of handling both simple queries, and ones requiring complex multi-hop retrieval and reasoning.

Our evaluation highlights the importance of specialized retrieval techniques for domain-specific applications and underscores the limitations of current LLM-based assessment metrics in financial contexts. Human evaluation proved crucial for accurately measuring performance, revealing substantial improvements with our ensembled approach. The modular design of our framework facilitates adaptation to other domains requiring precise information extraction. By addressing fundamental challenges in financial RAG systems, our work contributes to building more trustworthy AI systems for high-stakes applications where factual accuracy is paramount. Future research directions include fine-tuning models for financial contexts and developing more nuanced evaluation metrics.

\section*{Limitations}
Due to resource constraints, our evaluation is conducted on a relatively small dataset, which may limit the generalizability of the results.

Although our approach demonstrates improvements over existing baselines, its practical deployment is still challenged by the presence of hallucinations in more complex and ambiguous datasets. Consequently, the system currently requires human oversight and verification to ensure reliability and factual consistency.

\section*{Acknowledgements}
The authors thank InterIIT Tech Meet 13.0 and Pathway for proposing the problem statement and facilitating access to task materials and clarifications during the competition.

\bibliography{custom}

\begin{thebibliography}{42}
\providecommand{\natexlab}[1]{#1}

\bibitem[{Asai et~al.(2023)Asai, Wu, Wang, Sil, and Hajishirzi}]{asai2023selfraglearningretrievegenerate}
Akari Asai, Zeqiu Wu, Yizhong Wang, Avirup Sil, and Hannaneh Hajishirzi. 2023.
\newblock \href {https://doi.org/10.48550/arXiv.2310.11511} {Self-{{RAG}}: {{Learning}} to {{Retrieve}}, {{Generate}}, and {{Critique}} through {{Self-Reflection}}}.
\newblock \emph{Preprint}, arXiv:2310.11511.

\bibitem[{Chang et~al.(2024)Chang, Jiang, Rakesh, Pan, Yeh, Wang, Hu, Xu, Zheng, Das, and Zou}]{chang2024mainragmultiagentfilteringretrievalaugmented}
Chia-Yuan Chang, Zhimeng Jiang, Vineeth Rakesh, Menghai Pan, Chin-Chia~Michael Yeh, Guanchu Wang, Mingzhi Hu, Zhichao Xu, Yan Zheng, Mahashweta Das, and Na~Zou. 2024.
\newblock \href {https://doi.org/10.48550/arXiv.2501.00332} {{{MAIN-RAG}}: {{Multi-Agent Filtering Retrieval-Augmented Generation}}}.
\newblock \emph{Preprint}, arXiv:2501.00332.

\bibitem[{Chen et~al.(2022)Chen, Li, Smiley, Ma, Shah, and Wang}]{chen2022convfinqaexploringchainnumerical}
Zhiyu Chen, Shiyang Li, Charese Smiley, Zhiqiang Ma, Sameena Shah, and William~Yang Wang. 2022.
\newblock \href {https://doi.org/10.48550/arXiv.2210.03849} {{{ConvFinQA}}: {{Exploring}} the {{Chain}} of {{Numerical Reasoning}} in {{Conversational Finance Question Answering}}}.
\newblock \emph{Preprint}, arXiv:2210.03849.

\bibitem[{Chowdhery et~al.(2022)Chowdhery, Narang, Devlin, Bosma, Mishra, Roberts, Barham, Chung, Sutton, Gehrmann et~al.}]{chowdhery2022palm}
Aakanksha Chowdhery, Sharan Narang, Jacob Devlin, Maarten Bosma, Gaurav Mishra, Adam Roberts, Paul Barham, Hyung~Won Chung, Charles Sutton, Sebastian Gehrmann, and 1 others. 2022.
\newblock Palm: Scaling language modeling with pathways.
\newblock \emph{arXiv preprint arXiv:2204.02311}.

\bibitem[{Eibich et~al.(2024)Eibich, Nagpal, and {Fred-Ojala}}]{eibich2024aragogadvancedragoutput}
Matou{\v s} Eibich, Shivay Nagpal, and Alexander {Fred-Ojala}. 2024.
\newblock \href {https://doi.org/10.48550/arXiv.2404.01037} {{{ARAGOG}}: {{Advanced RAG Output Grading}}}.
\newblock \emph{Preprint}, arXiv:2404.01037.

\bibitem[{Es et~al.(2023)Es, James, {Espinosa-Anke}, and Schockaert}]{es2023ragasautomatedevaluationretrieval}
Shahul Es, Jithin James, Luis {Espinosa-Anke}, and Steven Schockaert. 2023.
\newblock \href {https://doi.org/10.48550/arXiv.2309.15217} {{{RAGAS}}: {{Automated Evaluation}} of {{Retrieval Augmented Generation}}}.
\newblock \emph{Preprint}, arXiv:2309.15217.

\bibitem[{Gao et~al.(2023)Gao, Ma, Lin, and Callan}]{gao2023precisezeroshotdenseretrieval}
Luyu Gao, Xueguang Ma, Jimmy Lin, and Jamie Callan. 2023.
\newblock \href {https://doi.org/10.18653/v1/2023.acl-long.99} {Precise {{Zero-Shot Dense Retrieval}} without {{Relevance Labels}}}.
\newblock In \emph{Proceedings of the 61st {{Annual Meeting}} of the {{Association}} for {{Computational Linguistics}} ({{Volume}} 1: {{Long Papers}})}, pages 1762--1777, Toronto, Canada. Association for Computational Linguistics.

\bibitem[{Gao et~al.(2021)Gao, Yao, and Chen}]{gao2021simcse}
Tianyu Gao, Xingcheng Yao, and Danqi Chen. 2021.
\newblock Simcse: Simple contrastive learning of sentence embeddings.
\newblock In \emph{Proceedings of the 2021 Conference on Empirical Methods in Natural Language Processing}, pages 6894--6910.

\bibitem[{Girhepuje et~al.(2024)Girhepuje, Sajeev, Jain, Sikder, Varma, George, Srinivasan, Kurup, Sinha, and Mondal}]{girhepuje2024regainsenchantintelligenttool}
Sahil Girhepuje, Siva~Sankar Sajeev, Purvam Jain, Arya Sikder, Adithya~Rama Varma, Ryan George, Akshay~Govind Srinivasan, Mahendra Kurup, Ashmit Sinha, and Sudip Mondal. 2024.
\newblock \href {https://doi.org/10.48550/arXiv.2401.15724} {{{RE-GAINS}} \& {{EnChAnT}}: {{Intelligent Tool Manipulation Systems For Enhanced Query Responses}}}.
\newblock \emph{Preprint}, arXiv:2401.15724.

\bibitem[{Guo et~al.(2024)Guo, Xia, Yu, Ao, and Huang}]{guo2024lightragsimplefastretrievalaugmented}
Zirui Guo, Lianghao Xia, Yanhua Yu, Tu~Ao, and Chao Huang. 2024.
\newblock \href {https://doi.org/10.48550/arXiv.2410.05779} {{{LightRAG}}: {{Simple}} and {{Fast Retrieval-Augmented Generation}}}.
\newblock \emph{Preprint}, arXiv:2410.05779.

\bibitem[{Guu et~al.(2020)Guu, Lee, Tung, Pasupat, and Chang}]{guu_realm_2020}
Kelvin Guu, Kenton Lee, Zora Tung, Panupong Pasupat, and Ming-Wei Chang. 2020.
\newblock \href {https://doi.org/10.48550/arXiv.2002.08909} {{REALM}: {Retrieval}-{Augmented} {Language} {Model} {Pre}-{Training}}.
\newblock \emph{arXiv preprint}.
\newblock ArXiv:2002.08909 [cs].

\bibitem[{Hao et~al.(2023)Hao, Gu, Ma, Hong, Wang, Wang, and Hu}]{hao2023reasoninglanguagemodelplanning}
Shibo Hao, Yi~Gu, Haodi Ma, Joshua~Jiahua Hong, Zhen Wang, Daisy~Zhe Wang, and Zhiting Hu. 2023.
\newblock \href {https://doi.org/10.48550/arXiv.2305.14992} {Reasoning with {{Language Model}} is {{Planning}} with {{World Model}}}.
\newblock \emph{Preprint}, arXiv:2305.14992.

\bibitem[{Henderson et~al.(2023)Henderson, Sinha, Angelard-Gontier, Ke, Fried, Lowe, and Pineau}]{henderson2023foundation}
Peter Henderson, Koustuv Sinha, Nicolas Angelard-Gontier, Nan~Rosemary Ke, Genevi{\`e}ve Fried, Ryan Lowe, and Joelle Pineau. 2023.
\newblock Foundation models for legal reasoning.
\newblock \emph{arXiv preprint arXiv:2307.03557}.

\bibitem[{Hu et~al.(2021)Hu, Shen, Wallis, {Allen-Zhu}, Li, Wang, Wang, and Chen}]{hu2021loralowrankadaptationlargeb}
Edward~J. Hu, Yelong Shen, Phillip Wallis, Zeyuan {Allen-Zhu}, Yuanzhi Li, Shean Wang, Lu~Wang, and Weizhu Chen. 2021.
\newblock \href {https://openreview.net/forum?id=nZeVKeeFYf9} {{{LoRA}}: {{Low-Rank Adaptation}} of {{Large Language Models}}}.
\newblock In \emph{International {{Conference}} on {{Learning Representations}}}.

\bibitem[{Huang et~al.(2023)Huang, Sun, Xiong, Dou, Zhang, and Yuan}]{huang2023survey}
Yue Huang, Xiaohan Sun, Yao Xiong, Zhicheng Dou, Guoliang Zhang, and Jian Yuan. 2023.
\newblock A survey on hallucination in large language models: Principles, taxonomy, challenges, and open questions.
\newblock \emph{arXiv preprint arXiv:2311.05232}.

\bibitem[{Islam et~al.(2023)Islam, Kannappan, Kiela, Qian, Scherrer, and Vidgen}]{islam2023financebenchnewbenchmarkfinancial}
Pranab Islam, Anand Kannappan, Douwe Kiela, Rebecca Qian, Nino Scherrer, and Bertie Vidgen. 2023.
\newblock \href {https://doi.org/10.48550/arXiv.2311.11944} {{{FinanceBench}}: {{A New Benchmark}} for {{Financial Question Answering}}}.
\newblock \emph{Preprint}, arXiv:2311.11944.

\bibitem[{Ji et~al.(2023)Ji, Lee, Frieske, Yu, Su, Xu, Ishii, Bang, Madotto, and Fung}]{ji2023survey}
Ziwei Ji, Nayeon Lee, Rita Frieske, Tiezheng Yu, Dan Su, Yan Xu, Etsuko Ishii, Ye~Jin Bang, Andrea Madotto, and Pascale Fung. 2023.
\newblock Survey of hallucination in natural language generation.
\newblock \emph{ACM Computing Surveys}.

\bibitem[{Karpukhin et~al.(2020)Karpukhin, Oguz, Min, Lewis, Wu, Edunov, Chen, and Yih}]{karpukhin2020dense}
Vladimir Karpukhin, Barlas Oguz, Sewon Min, Patrick Lewis, Ledell Wu, Sergey Edunov, Danqi Chen, and Wen-tau Yih. 2020.
\newblock Dense passage retrieval for open-domain question answering.
\newblock \emph{arXiv preprint arXiv:2004.04906}.

\bibitem[{Khattab and Zaharia(2020)}]{khattab2020colbert}
Omar Khattab and Matei Zaharia. 2020.
\newblock Colbert: Efficient and effective passage search via contextualized late interaction over bert.
\newblock \emph{Proceedings of the 43rd International ACM SIGIR conference on research and development in Information Retrieval}, pages 39--48.

\bibitem[{{LangChain}(2023)}]{langchain2023querytransformations}
{LangChain}. 2023.
\newblock \href {https://blog.langchain.com/query-transformations/} {Query {{Transformations}}}.

\bibitem[{Lewis et~al.(2020)Lewis, Perez, Piktus, Petroni, Karpukhin, Goyal, Küttler, Lewis, Yih, Rocktäschel et~al.}]{lewis2020retrieval}
Patrick Lewis, Ethan Perez, Aleksandra Piktus, Fabio Petroni, Vladimir Karpukhin, Naman Goyal, Heinrich Küttler, Mike Lewis, Wen-tau Yih, Tim Rocktäschel, and 1 others. 2020.
\newblock Retrieval-augmented generation for knowledge-intensive nlp tasks.
\newblock \emph{Advances in Neural Information Processing Systems}, 33:9459--9474.

\bibitem[{Li et~al.(2024)Li, Wang, Jiang, Mao, Chen, Du, Zhang, Zhang, Zhang, and Liu}]{li2024dmqrragdiversemultiqueryrewriting}
Zhicong Li, Jiahao Wang, Zhishu Jiang, Hangyu Mao, Zhongxia Chen, Jiazhen Du, Yuanxing Zhang, Fuzheng Zhang, Di~Zhang, and Yong Liu. 2024.
\newblock \href {https://arxiv.org/abs/2411.13154} {Dmqr-rag: Diverse multi-query rewriting for rag}.
\newblock \emph{Preprint}, arXiv:2411.13154.

\bibitem[{Li et~al.(2023)Li, Wang, Chen, and Chen}]{li2023finbert}
Zhuangzhuang Li, Hanyi Wang, Zhengqing Chen, and Xia Chen. 2023.
\newblock Finbert: A pre-trained financial language representation model for financial text mining.
\newblock \emph{Proceedings of the Twenty-Second International Joint Conference on Artificial Intelligence}.

\bibitem[{Liu et~al.(2023)Liu, Xie, Chen, Wang, Yuan, Liu, Hu, Wang, Qiao, Pan et~al.}]{liu2023toolllm}
Yujia Liu, Yuwei Xie, Chunyuan Chen, Sylvia Wang, Yuxin Yuan, Yang Liu, Xiang Hu, Songyang Wang, Tianyu Qiao, Lingyu Pan, and 1 others. 2023.
\newblock Toolllm: Facilitating large language models to master 16000+ real-world apis.
\newblock \emph{arXiv preprint arXiv:2307.16789}.

\bibitem[{OpenAI et~al.(2024)OpenAI, Achiam, Adler, Agarwal, Ahmad, Akkaya, Aleman, Almeida, Altenschmidt, Altman, Anadkat, Avila, Babuschkin, Balaji, Balcom, Baltescu, Bao, Bavarian, Belgum, Bello, Berdine, Bernadett-Shapiro, Berner, Bogdonoff, Boiko, Boyd, Brakman, Brockman, Brooks, Brundage, Button, Cai, Campbell, Cann, Carey, Carlson, Carmichael, Chan, Chang, Chantzis, Chen, Chen, Chen, Chen, Chen, Chess, Cho, Chu, Chung, Cummings, Currier, Dai, Decareaux, Degry, Deutsch, Deville, Dhar, Dohan, Dowling, Dunning, Ecoffet, Eleti, Eloundou, Farhi, Fedus, Felix, Fishman, Forte, Fulford, Gao, Georges, Gibson, Goel, Gogineni, Goh, Gontijo-Lopes, Gordon, Grafstein, Gray, Greene, Gross, Gu, Guo, Hallacy, Han, Harris, He, Heaton, Heidecke, Hesse, Hickey, Hickey, Hoeschele, Houghton, Hsu, Hu, Hu, Huizinga, Jain, Jain, Jang, Jiang, Jiang, Jin, Jin, Jomoto, Jonn, Jun, Kaftan, Łukasz Kaiser, Kamali, Kanitscheider, Keskar, Khan, Kilpatrick, Kim, Kim, Kim, Kirchner, Kiros, Knight, Kokotajlo, Łukasz Kondraciuk,
  Kondrich, Konstantinidis, Kosic, Krueger, Kuo, Lampe, Lan, Lee, Leike, Leung, Levy, Li, Lim, Lin, Lin, Litwin, Lopez, Lowe, Lue, Makanju, Malfacini, Manning, Markov, Markovski, Martin, Mayer, Mayne, McGrew, McKinney, McLeavey, McMillan, McNeil, Medina, Mehta, Menick, Metz, Mishchenko, Mishkin, Monaco, Morikawa, Mossing, Mu, Murati, Murk, Mély, Nair, Nakano, Nayak, Neelakantan, Ngo, Noh, Ouyang, O'Keefe, Pachocki, Paino, Palermo, Pantuliano, Parascandolo, Parish, Parparita, Passos, Pavlov, Peng, Perelman, de~Avila Belbute~Peres, Petrov, de~Oliveira~Pinto, Michael, Pokorny, Pokrass, Pong, Powell, Power, Power, Proehl, Puri, Radford, Rae, Ramesh, Raymond, Real, Rimbach, Ross, Rotsted, Roussez, Ryder, Saltarelli, Sanders, Santurkar, Sastry, Schmidt, Schnurr, Schulman, Selsam, Sheppard, Sherbakov, Shieh, Shoker, Shyam, Sidor, Sigler, Simens, Sitkin, Slama, Sohl, Sokolowsky, Song, Staudacher, Such, Summers, Sutskever, Tang, Tezak, Thompson, Tillet, Tootoonchian, Tseng, Tuggle, Turley, Tworek, Uribe, Vallone,
  Vijayvergiya, Voss, Wainwright, Wang, Wang, Wang, Ward, Wei, Weinmann, Welihinda, Welinder, Weng, Weng, Wiethoff, Willner, Winter, Wolrich, Wong, Workman, Wu, Wu, Wu, Xiao, Xu, Yoo, Yu, Yuan, Zaremba, Zellers, Zhang, Zhang, Zhao, Zheng, Zhuang, Zhuk, and Zoph}]{openai2024gpt4technicalreport}
OpenAI, Josh Achiam, Steven Adler, Sandhini Agarwal, Lama Ahmad, Ilge Akkaya, Florencia~Leoni Aleman, Diogo Almeida, Janko Altenschmidt, Sam Altman, Shyamal Anadkat, Red Avila, Igor Babuschkin, Suchir Balaji, Valerie Balcom, Paul Baltescu, Haiming Bao, Mohammad Bavarian, Jeff Belgum, and 262 others. 2024.
\newblock \href {https://arxiv.org/abs/2303.08774} {Gpt-4 technical report}.
\newblock \emph{Preprint}, arXiv:2303.08774.

\bibitem[{Qin et~al.(2023)Qin, Deng, Xu, Chen, Lin, Sun, Bu, Li, Zhou, Yang et~al.}]{qin2023tool}
Yujia Qin, Shengding Deng, Furui Xu, Shiwei Chen, Yankai Lin, Weilin Sun, Meng Bu, Peng Li, Shulin Zhou, Chao Yang, and 1 others. 2023.
\newblock Tool learning with foundation models.
\newblock \emph{arXiv preprint arXiv:2304.08354}.

\bibitem[{Radhakrishnan et~al.(2023)Radhakrishnan, Nguyen, Chen, Chen, Denison, Hernandez, Durmus, Hubinger, Kernion, Lukošiūtė, Cheng, Joseph, Schiefer, Rausch, McCandlish, Showk, Lanham, Maxwell, Chandrasekaran, Hatfield-Dodds, Kaplan, Brauner, Bowman, and Perez}]{radhakrishnan2023questiondecompositionimprovesfaithfulness}
Ansh Radhakrishnan, Karina Nguyen, Anna Chen, Carol Chen, Carson Denison, Danny Hernandez, Esin Durmus, Evan Hubinger, Jackson Kernion, Kamilė Lukošiūtė, Newton Cheng, Nicholas Joseph, Nicholas Schiefer, Oliver Rausch, Sam McCandlish, Sheer~El Showk, Tamera Lanham, Tim Maxwell, Venkatesa Chandrasekaran, and 5 others. 2023.
\newblock \href {https://arxiv.org/abs/2307.11768} {Question decomposition improves the faithfulness of model-generated reasoning}.
\newblock \emph{Preprint}, arXiv:2307.11768.

\bibitem[{Reimers and Gurevych(2019)}]{reimers2019sentence}
Nils Reimers and Iryna Gurevych. 2019.
\newblock Sentence-bert: Sentence embeddings using siamese bert-networks.
\newblock In \emph{Proceedings of the 2019 Conference on Empirical Methods in Natural Language Processing and the 9th International Joint Conference on Natural Language Processing (EMNLP-IJCNLP)}, pages 3982--3992.

\bibitem[{Schick et~al.(2023)Schick, Dwivedi-Yu, Dess{\`i}, Raileanu, Lomeli, Zettlemoyer, Cancedda, and Scialom}]{schick2023toolformer}
Timo Schick, Jane Dwivedi-Yu, Roberto Dess{\`i}, Roberta Raileanu, Maria Lomeli, Luke Zettlemoyer, Nicola Cancedda, and Thomas Scialom. 2023.
\newblock Toolformer: Language models can teach themselves to use tools.
\newblock \emph{arXiv preprint arXiv:2302.04761}.

\bibitem[{Singhal et~al.(2023)Singhal, Azizi, Tu, Mahdavi, Wei, Chung, Scales, Venkataraman, Maginnis, Nori et~al.}]{singhal2023large}
Karan Singhal, Shekoofeh Azizi, Tao Tu, S~Sara Mahdavi, Jason Wei, Hyung~Won Chung, Nathan Scales, Ajay Venkataraman, Gabriel Maginnis, Arun Nori, and 1 others. 2023.
\newblock Large language models in medicine.
\newblock \emph{Nature Medicine}, 29(8):1998--2012.

\bibitem[{Touvron et~al.(2023)Touvron, Lavril, Izacard, Martinet, Lachaux, Lacroix, Rozière, Goyal, Hambro, Azhar et~al.}]{touvron2023llama}
Hugo Touvron, Thibaut Lavril, Gautier Izacard, Xavier Martinet, Marie-Anne Lachaux, Timothée Lacroix, Baptiste Rozière, Naman Goyal, Eric Hambro, Faisal Azhar, and 1 others. 2023.
\newblock Llama: Open and efficient foundation language models.
\newblock \emph{arXiv preprint arXiv:2302.13971}.

\bibitem[{Wang et~al.(2023)Wang, Xu, Lan, Hu, Lan, Lee, and Lim}]{wang2023planandsolvepromptingimprovingzeroshot}
Lei Wang, Wanyu Xu, Yihuai Lan, Zhiqiang Hu, Yunshi Lan, Roy Ka-Wei Lee, and Ee-Peng Lim. 2023.
\newblock \href {https://arxiv.org/abs/2305.04091} {Plan-and-solve prompting: Improving zero-shot chain-of-thought reasoning by large language models}.
\newblock \emph{Preprint}, arXiv:2305.04091.

\bibitem[{Wang et~al.(2025)Wang, Chi, Tai, Kwok, Li, Li, He, Hua, Lu, Wang, Wu, Huang, Tian, Mo, Cui, and Zhou}]{wang2025finsagemultiaspectragfinancial}
Xinyu Wang, Jijun Chi, Zhenghan Tai, Tung Sum~Thomas Kwok, Muzhi Li, Zhuhong Li, Hailin He, Yuchen Hua, Peng Lu, Suyuchen Wang, Yihong Wu, Jerry Huang, Jingrui Tian, Fengran Mo, Yufei Cui, and Ling Zhou. 2025.
\newblock \href {https://arxiv.org/abs/2504.14493} {Finsage: A multi-aspect rag system for financial filings question answering}.
\newblock \emph{Preprint}, arXiv:2504.14493.

\bibitem[{Wu et~al.(2023)Wu, Irsoy, Lu, Dabravolski, Dredze, Gehrmann, Kambadur, Rosenberg, and Mann}]{wu2023bloomberggpt}
Shijie Wu, Ozan Irsoy, Steven Lu, Vadim Dabravolski, Mark Dredze, Sebastian Gehrmann, Prabhanjan Kambadur, David Rosenberg, and Gideon Mann. 2023.
\newblock Bloomberggpt: A large language model for finance.
\newblock \emph{arXiv preprint arXiv:2303.17564}.

\bibitem[{Wu et~al.(2024)Wu, Yue, Zhang, Wang, and Wu}]{wu2024stateflowenhancingllmtasksolving}
Yiran Wu, Tianwei Yue, Shaokun Zhang, Chi Wang, and Qingyun Wu. 2024.
\newblock \href {https://doi.org/10.48550/arXiv.2403.11322} {{{StateFlow}}: {{Enhancing LLM Task-Solving}} through {{State-Driven Workflows}}}.
\newblock \emph{Preprint}, arXiv:2403.11322.

\bibitem[{Xiong et~al.(2020)Xiong, Xiong, Li, Tang, Liu, Bennett, Ahmed, and Overwijk}]{xiong2020approximate}
Lee Xiong, Chenyan Xiong, Ye~Li, Kwok-Fung Tang, Jialin Liu, Paul Bennett, Junaid Ahmed, and Arnold Overwijk. 2020.
\newblock Approximate nearest neighbor negative contrastive learning for dense text retrieval.
\newblock \emph{arXiv preprint arXiv:2007.00808}.

\bibitem[{Yan et~al.(2024)Yan, Gu, Zhu, and Ling}]{yan2024correctiveretrievalaugmentedgeneration}
Shi-Qi Yan, Jia-Chen Gu, Yun Zhu, and Zhen-Hua Ling. 2024.
\newblock \href {https://doi.org/10.48550/arXiv.2401.15884} {Corrective {{Retrieval Augmented Generation}}}.
\newblock \emph{Preprint}, arXiv:2401.15884.

\bibitem[{Yang et~al.(2024)Yang, Zhang, Wang, Guo, Zhang, Lin, Wang, Zhou, Guan, Zhang, and Wang}]{yang2024finrobotopensourceaiagent}
Hongyang Yang, Boyu Zhang, Neng Wang, Cheng Guo, Xiaoli Zhang, Likun Lin, Junlin Wang, Tianyu Zhou, Mao Guan, Runjia Zhang, and Christina~Dan Wang. 2024.
\newblock \href {https://doi.org/10.48550/arXiv.2405.14767} {{{FinRobot}}: {{An Open-Source AI Agent Platform}} for {{Financial Applications}} using {{Large Language Models}}}.
\newblock \emph{Preprint}, arXiv:2405.14767.

\bibitem[{Yao et~al.(2022)Yao, Zhao, Yu, Du, Shafran, Narasimhan, and Cao}]{yao2022react}
Shunyu Yao, Jeffrey Zhao, Dian Yu, Nan Du, Izhak Shafran, Karthik Narasimhan, and Yuan Cao. 2022.
\newblock React: Synergizing reasoning and acting in language models.
\newblock \emph{arXiv preprint arXiv:2210.03629}.

\bibitem[{Zhang et~al.(2022)Zhang, Shlens, and Dean}]{zhang2022designing}
Barret Zhang, Jonathon Shlens, and Jeff Dean. 2022.
\newblock Designing effective sparse expert models.
\newblock \emph{arXiv preprint arXiv:2202.08906}.

\bibitem[{Zhang et~al.(2024)Zhang, Wu, Yang, and Nie}]{zhang2024exploringbestpracticesquery}
Le~Zhang, Yihong Wu, Qian Yang, and Jian-Yun Nie. 2024.
\newblock \href {https://arxiv.org/abs/2401.06311} {Exploring the best practices of query expansion with large language models}.
\newblock \emph{Preprint}, arXiv:2401.06311.

\bibitem[{Zhou et~al.(2023)Zhou, Schärli, Hou, Wei, Scales, Wang, Schuurmans, Cui, Bousquet, Le, and Chi}]{zhou2023leasttomostpromptingenablescomplex}
Denny Zhou, Nathanael Schärli, Le~Hou, Jason Wei, Nathan Scales, Xuezhi Wang, Dale Schuurmans, Claire Cui, Olivier Bousquet, Quoc Le, and Ed~Chi. 2023.
\newblock \href {https://arxiv.org/abs/2205.10625} {Least-to-most prompting enables complex reasoning in large language models}.
\newblock \emph{Preprint}, arXiv:2205.10625.

\end{thebibliography}

\appendix

\section{Question-answer example}
\label{json-example}

\begin{minted}[breaklines=true]{json}
{     
    "question": "For American Water Works, what was the rate of growth from 2013 to 2014 in the fair value per share"
    "answer" : "",
    "context": "~13.3%. Page 81, Table[The 
    weighted-average assumptions used in the 
    Monte Carlo simulation and the weighted-average 
    grant date fair values of RSUs granted 
    for the years ended December 31]
    
    [45.45 -40.13]/40.13 = 13.3%
    
    AMERICAN WATER WORKS COMPANY, INC.",
    "ticker": "AWK",
    "filed on": "31 December 2015"
}
\end{minted}
\section{Meta-Plan JSON Instructions}
\label{sec:metaplanjsonscehma}

\begin{minted}[breaklines=true]{python}
{
    "thought": "...",  # Thought process and reasoning of the bot for the current step
    "tool_calls": [{"name": "...", "args": {...}}, 
    {"name": "...", "args": {...}}, ...],  # List of tools to be called along with the appropriate arguments. 
    "audio": "...", # Respond comprehensively to the query in a verbose way and output in formatted markdown string
    "plan": "...", 
    # The overall plan for calling various tools and answering the query. This needs to be updated dynamically based on the retrieved information from tool calls.
    "queries": [{"query":"...","answer":"..."},
    {"query":"...","answer":"..."}] 
    # The decomposed sub-queries. Intially all the answers are empty strings, as answers from tool calls come in, update them accordingly
}

\end{minted}
\section{Retrieval Challenges} \label{retrieval-challenges}

In the financial domain, retrieval methods that rely solely on vector similarity often fail to distinguish between passages that are semantically similar but differ in critical numerical details or temporal references. These distinctions, although subtle, are essential for producing accurate and trustworthy responses when answering questions about structured financial reports.

Consider the following example from our evaluation set:

\begin{minted}[breaklines=true]{js} 
"query": "For American Water Works, what was the growth in allowance for other funds used during construction from 2013 to 2014?"

"retrieved_reference_1": "In 2014, we spent $3.6 million, including $0.8 million funded by research grants... (discussion on research and development spending) [awk_2015_10K.pdf]"

"retrieved_reference_2": "Cash flows used in investing activities increased in 2014 compared to 2013 primarily due to an increase in our capital expenditures... (details on capital expenditures) [awk_2015_10K.pdf]"

"retrieved_reference_3": "Amortization of contributions in aid of construction was $23,913, $22,363, and $20,979 for the years ended December 31, 2014, 2013, and 2012... (amortization details) [awk_2015_10K.pdf]"

"retrieved_reference_4": "Such grants reduce the cost of research and allow collaboration with leading national and international researchers... (discussion on research grants and collaboration) [awk_2017_10K.pdf]"

"retrieved_reference_5": "Amortization of contributions in aid of construction was $27, $26, and $24 for the years ended December 31, 2016, 2015, and 2014... (further amortization details) [awk_2017_10K.pdf]"

"retrieved_reference_6": "The estimated capital expenditures required under legal and binding contractual obligations amounted to $217,082 at December 31, 2014... (details on capital expenditure commitments) [awk_2015_10K.pdf]" \end{minted}

As shown above, SEC filings from different years (e.g., 2015 vs. 2017) often include passages with similar or even nearly identical phrasing. However, for financial question-answering, distinctions such as the reporting year or specific numerical values are vital for correctness. Standard dense retrieval models tend to conflate these passages due to their semantic resemblance, leading to unreliable results.

To mitigate this issue, we incorporate BM25 alongside dense vector retrieval. This hybrid approach ensures that keyword and phrase-level matches (e.g., exact years, financial figures, or domain-specific terminology) complement semantic similarity, resulting in more precise and contextually appropriate retrievals.
\section{Tools}
\label{subsec:toolset}
Our agentic pipeline has various tools to fetch data from various data sources apart from the retrieved context. The tools are divided into different types based on their use cases give below. Having more than one tools provide redundancy in case one or more tools fail.

\begin{enumerate}
    \item \textbf{Web-search: } The web search tool provides real-time access to the web search queries providing access to news, web pages, and more which might not be there in the retrieved context. \textbf{SERP API}, \textbf{Bing Web Search}, and \textbf{DuckDuckGo Web Search} are the tools used by the agent to obtain the data from a web search.

    \item \textbf{Financial Data API: } This is a collection of tools that provide real-time as well as historical data about the prices of stocks, securities, and cryptocurrencies. \textbf{Yahoo Finance}, \textbf{Alpha Vantage}, \textbf{EDGAR Tool}(Electronic Data Gathering, Analysis, and Retrieval system) and  \textbf{Financial Modelling Prep} tool providing real time financial data from various exchanges.

    \item \textbf{Mathematical tools: } The \textbf{WolframAlpha API} and \textbf{Python Calculator} are the tools incorporated to provide the the data processing ability to the agent. WolframAlpha takes in the mathematical questions in natural language and provides us with the answer whereas python calculator can be used to help the agent with more menial calculations.
    
\end{enumerate}

\section{Dataset}
\label{appendix: dataset}
Owing to the inconsistent evaluation results often observed in LLM-based methods and limited computational resources, we conduct our experiments on a focused subset of the FinanceBench and ConvFinQA datasets. Specifically, we select reports with the highest density of associated questions to ensure the relevance and informativeness of our evaluation. The selected subset comprises SEC 10-K filings from the following companies:

\begin{itemize} \item \textbf{American Water Works}: 2015, 2017, 2018 \item \textbf{AMD}: 2022 \item \textbf{American Express}: 2022 \item \textbf{Boeing}: 2022 \end{itemize}
\section{Evaluation}
\label{appendix:eval}
\subsection{Definitions of metrics}
\label{appendix:def-metrics}
RAGAS defines metrics by comparing the facts in a model's answer to those in the retrieved context or ground truth. The \emph{Faithfulness Score} is RAG-specific, measuring the proportion of claims in the answer that are supported by the retrieved context. \emph{Factual Correctness}, based on the F1 score, can be applied to any model.

\noindent\textbf{Faithfulness:}
\begin{equation}
\text{Faithfulness} =
\frac{\mathrm{Supported\ claims}}{\mathrm{Total\ claims\ in\ answer}}
\end{equation}

\noindent\textbf{Factual Correctness:}  
Let  
\begin{align*}
\text{TP} &= \#\{\text{claims in answer present in reference}\} \\
\text{FP} &= \#\{\text{claims in answer not in reference}\} \\
\text{FN} &= \#\{\text{claims in reference not in answer}\}
\end{align*}
Then
\begin{align}
\text{Precision} &= \frac{\text{TP}}{\text{TP}+\text{FP}},\\
\text{Recall} &= \frac{\text{TP}}{\text{TP}+\text{FN}}, \\
\text{F1} &= \frac{2\circ\text{Precision}\circ\text{Recall}}{\text{Precision}+\text{Recall}}
\end{align}
For human evalution, we define accuracy ($A$) as the fraction of correct answers\footnote{In cases when a question requires multiple independent answers, we assign the score as the fraction of correct answers for that question} and reliability ($R$) as the fraction of answers where the LLM was confident about its answer being correct. This is related to the hallucination rate (the fraction of incorrect answers, denoted $HR$) as
\begin{equation}
    HR = \frac{A}{R}-A, \text{where}
\end{equation}
    
\begin{align*}
U &= \text{Number of times LLM refused to answer} \\
N &= \text{total number of answers}
\end{align*}

In Table \ref{tab: human_eval}, we rely on human evaluation to show that the Agentic pipeline performs better than Multi-HyDE with tools.

\subsection{Examples in which RAGAS fails}
\label{appendix:ragas-failure}
In addition to errors in the LLM evaluation used by RAGAS, we note that the evaluation often fails when more precise numbers are given in the generated answer than available in the ground truth. This is partly due to ambiguities in the dataset, but also because, for instance, it often penalizes the model for minor inaccuracies or rounding, and at other times it gives a high score despite an outright refusal to answer.

\lstset{
  breaklines=true,
  frame=single,
}
\begin{lstlisting}[caption={RAGAS incorrectly gives a factual correctness of 0 though the response is simply more precise than the reference.}]
Question: For American Water Works, what was the rate of growth from 2013 to 2014 in the fair value per share
Response: ```json
{
  "tool_calls": []
}
```

Based on the provided text, the fair value per share for American Water Works was $37.40 in 2012, $40.13 in 2013, and $45.45 in 2014 (Document 8).  Therefore, the growth rate from 2013 to 2014 is:

($45.45 - $40.13) / $40.13 = 0.1325 or 13.25%



Reference: 13.30%
Semantic Similarity: 0.5918684811439618
Factual Correctness: 0.0
Faithfulness: 0.6666666666666666
\end{lstlisting}

\begin{lstlisting}[caption={RAGAS incorrectly scores the LLM's refusal to answer very highly}]
Question: For American Water Works, what was the rate of growth from 2013 to 2014 in the fair value per share
Response: The provided text does not contain information on the fair value per share of American Water Works for 2013 and 2014, nor does it provide the means to calculate a growth rate.  Therefore, I cannot answer your question using the given documents.  To answer this question, additional information is needed.  I would need to access financial statements or other reliable sources containing this data.

Reference: 13.30%
Semantic Similarity: 0.5736175582900219
Factual Correctness: 0.91
Faithfulness: 1.0

\end{lstlisting}
\section{Additional Implementation Details}
\label{implementation-details}
\begin{figure}[htbp]
    \centering
    \includegraphics[width=1\linewidth]{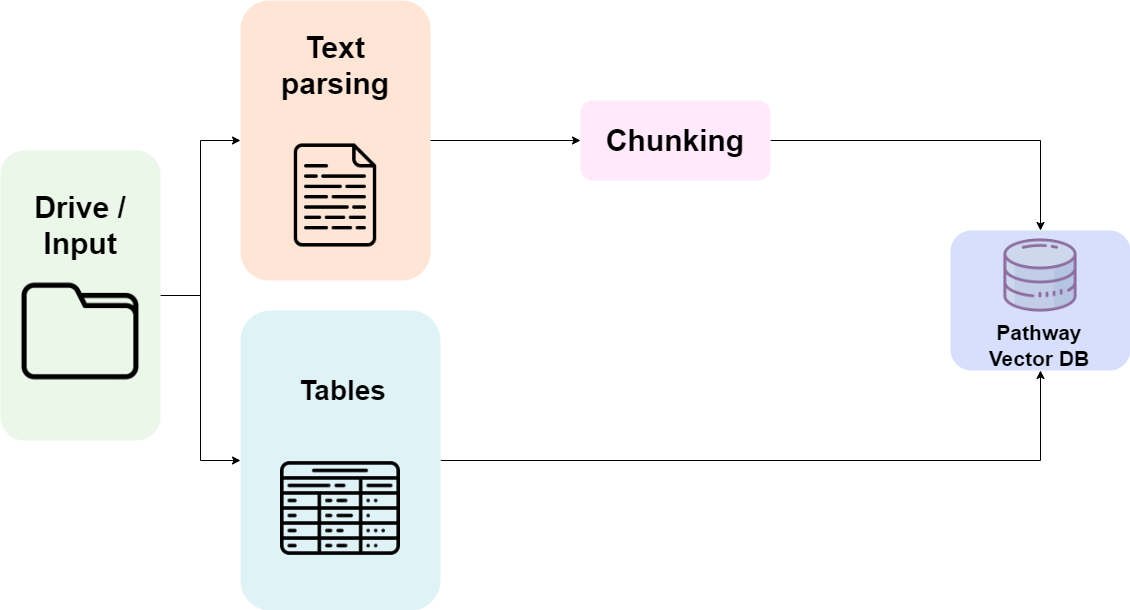}
    \caption{PDF-Parser System: Documents can be added to a Google Drive, which dynamically updates the Vector Store. Text is parsed using our parsing pipeline, chunked recursively before populating the Vector Store.}
    \label{fig:parser}
\end{figure}

\paragraph{PDF Parsing}
We use a customised PDF parsing system (Fig. \ref{fig:parser}) using the \textbf{Docling} library to extract and structure data from complex documents. It handles text, tables, and images, exporting tables in HTML format. Further, it utilizes recursive chunk splitting as the text chunking strategy for context preservation. 

\paragraph{Retrieval}
We employ \textbf{HNSW} for indexing and in addition use \textbf{BM25} for retrieving tables. Our Vector Store is implemented using Pathway \footnote{\url{https://github.com/pathwaycom/pathway}}. Row and column aggregation is also performed on tables. Keeping modularity in focus, retrieval methods are represented as tools, alongside others like web search and calculator.
The \textbf{Multi-HyDE} retriever, selects the top \( K_1 = 10 \) chunks, while the BM25 retriever fetches the top \( K_2 = 15 \) chunks.

\paragraph{Reranking} A re-ranker\footnote{\url{https://huggingface.co/cross-encoder/ms-marco-MiniLM-L-6-v2} or \url{https://huggingface.co/yxzwayne/bge-reranker-v2-m3}, specified in our experiments} is employed to pick the top \( K = 8 \) relevant chunks. This was determined after evaluating performance on various values of \( K \), as shown in Table \ref{tab: accuracy-for-different-k}.\\
\begin{table}[htbp]
    \centering
    \begin{tabular}{@{}cc@{}}
        \toprule
        \textbf{Top K Value} & \textbf{Accuracy (\%)} \\
        \midrule
        1 & 57.5 \\
        2 & 75.3 \\
        8 & 79.6 \\
        10 & \textbf{80.1} \\
        \bottomrule
    \end{tabular}
    \caption{Accuracy for different values of 10 K retrieved documents.}
    \label{tab: accuracy-for-different-k}
\end{table}
 \clearpage

\section{Other Ablations}
Tables below depict other ablations performed as a part of our experimentation and analysis.
\label{appendix: ablations}
\begin{table}[htb]
    \centering
    \begin{tabular}{@{}lccc@{}}
        \toprule
        \textbf{Method} & \textbf{Precision} & \textbf{Recall} & \textbf{Accuracy} \\
        \midrule
        Naive-RAG & 0.912 & 0.592 & 0.616 \\
        HyDE & TBD & TBD & TBD \\
        Multi-HyDE & 0.932 & 0.625 & 0.721 \\
        \bottomrule
    \end{tabular}
    \caption{Comparison of Naive-RAG, HyDE, and Multi-HyDE on a subset of financial-qa-10K dataset}
    \label{tab:retrieval-comparison}
\end{table}
\begin{table}[htb]
    \centering
    \begin{tabular}{@{}lccc@{}}
        \toprule
        \textbf{Method} & \textbf{In-Tokens} & \textbf{Out-Tokens} & \textbf{Time Spent} \\
        \midrule
        Naive-RAG & - & - & 0.199s \\
        HyDE & 133.5 & 428.2 & 9.344s \\
        Multi-HyDE & 193.6 & 421.4 & 9.121s \\
        \bottomrule
    \end{tabular}
    \caption{Resource usage comparison on Financial-qa-10K Dataset.}
    \label{tab:resource-comparison}
\end{table}

\begin{table}[h]
    \centering
    \begin{adjustbox}{max width=1\textwidth}
    
    \begin{tabular}{@{}cccc@{}}
        \toprule
        \textbf{Metric} & \textbf{Method 1} & \textbf{Method 2} \\
        \midrule
        Cosine Similarity & \textbf{0.5981} & 0.5765 \\
        Recall & 0.2462 & \textbf{0.2910} \\
        Factual Correctness & 0.1738 & \textbf{0.2291} \\
        Faithfulness & 0.8103 & \textbf{0.8754} \\
        ROUGE score & 0.0346 & \textbf{0.0349} \\
        \bottomrule
    \end{tabular}
    \end{adjustbox}
    \caption{Evaluation metrics comparing different parsers, showing the improvement of Docling over Open Parse (with ConvFinQA and FinBench dataset subsets) \\
    \\
    Method 1: Multi-HyDE + Re-ranker + Open Parse (Llama-70b) \\
    Method 2: Multi-HyDE + Re-ranker + Docling Parser (Llama-70b) \\}
    \label{tab: evaluation metrics comparing parsers (transposed)}
\end{table}

\end{document}